\documentstyle[prl,aps]{revtex}
\begin{document}


\title{On a Dual Standard Model\footnote{Based on talk given
at Solitons `97, Kingston, ON, Canada.}}

\author{Tanmay Vachaspati}
\address{
Physics Department,
Case Western Reserve University,
Cleveland OH 44106-7079, USA.}

\maketitle

\begin{abstract}

I describe various aspects of the construction of a dual 
standard model including how it may be possible to obtain the charge 
spectrum, the family structure and spin of the known matter particles.
I summarize the encouraging features of the model, the open
problems and the predictions indicated at this stage.

\end{abstract}

\section{Motivation}

The idea that particles in one theory are solitons of another theory
is best illustrated by the example of the equivalence of the  Sine-Gordon 
and Thirring model \cite{sc,sm}. The Sine-Gordon model contains both 
(scalar) particles
and solitons. At strong coupling, the model is better described by the
Thirring model whose fermionic particle excitations correspond to the
solitons of the Sine-Gordon model. If this duality is applicable to
the standard model as well, we might view the standard model as an
analog of the Thirring
model which then naturally leads us to consider what the dual standard
model (analog of the Sine-Gordon model) might be. 

Another motivation for the construction of a dual standard model is the
same motivation that has been with us from the turn of the century,
leading to efforts by eminent physicists to construct a model in which the 
particles we know (such as the electron) emerge as classical objects, or, in
modern language, as solitons. The most successful of these efforts is the
attempt by Skyrme \cite{skyrme1,skyrme2} that has led to a model for baryons 
and mesons about which we have heard so much at this conference. However, 
Skyrme did his historic work at a time when quarks had not been 
discovered. In this age, we would like to construct a Skyrme model for 
quarks and leptons.

Another motivating factor is that it may be simpler to
understand confinement in a dualized version of QCD where the
chromoelectric flux tubes appear as Nielsen-Olesen vortices \cite{hnpo}.
But this picture is incomplete since the quarks, on which the flux tubes are
supposed to terminate, have not yet been accounted for. From
studying the topological aspects of field theories, it is known that 
flux tubes can terminate on magnetic monopoles and so 
a more complete
dual superconductor picture of QCD would be one where quarks
correspond to magnetic monopoles which are confined by flux tubes
in color singlet combinations. To begin to paint such a picture,
the first step is to construct a model which gives rise to a spectrum
of magnetic monopoles which can be identified with the various
quarks we know.

\section{Construction of the Dual Model}

The simplest (toy)
version of the dual standard model we consider is described
by the ${\tilde {SU}}(5)$ invariant Lagrangian density
$$
L = - {1\over 4} F_{\mu \nu}^a F^{\mu \nu a} + |D\Phi |^2 - V(\Phi )\ ,
$$
where $\Phi$ is an adjoint of $\tilde {SU}(5)$ and the symbols have their
standard meanings. (Tildes on symmetry groups denote that they are the
dual symmetry groups and should not be confused with the symmetry
groups of the standard model.)
With some mild restrictions on the parameters
occurring in the quartic potential $V(\Phi )$, $\Phi$ gets a vacuum
expectation value and leads to the spontaneous symmetry breaking
$$
{\tilde {SU}} (5) \rightarrow
[ {\tilde {SU}} (3)\times {\tilde {SU}} (2) \times {\tilde {U}} (1) ]/
(Z_3 \times Z_2) \ .
$$
Note the discrete factors that need to be quotiented out. These are the
centers of ${\tilde {SU}} (3)$ and ${\tilde {SU}}(2)$ and also appear as
a $Z_3 \times Z_2$ subgroup of ${\tilde {U}} (1)$. The monopoles in this
model can be classified by their winding number which is an integer $n$.
However, again with some mild restrictions on the parameters of the Higgs
potential, we find that only the $n=\pm 1, \pm 2, \pm 3, \pm 4$, 
$\pm 6$ monopoles are stable. All other monopoles are unstable \cite{gh}. 
It is quite remarkable that these stable monopoles are in one to one 
correspondence with one family of standard model fermions \cite{tv,hltv}
(see Table I). 

\begin{center}
TABLE I
\small
Charges on classically stable ${\tilde {SU}}(5)$ monopoles and
the corresponding standard model fermions. For example,
$(u,d)_L$ have electric charges which are the same as
the magnetic charges on the $n=1$ monopole. Also shown 
are the monopole degeneracies $d_m$ which also correspond 
to the number of standard model fermions. 
\end{center}
\normalsize
\begin{center}
\begin{tabular*}{10.0cm}{|c@{\extracolsep{\fill}}cccc|c|}
\hline
&&&&&\\[-0.15cm]
                  {$n_{~}$}
                 & {$n_3$}
                 & {$n_2$}
                 & {$n_1$}
                 & {$d_m$}
                 & {$$}
                                   \\[0.20cm]
\hline
&&&&&\\[-0.1cm]
+1&1/3&1/2&+1/6&6&$(u,d)_L   $         \\[0.5cm]
-2&1/3&0  &-1/3&3&$d_R       $         \\[0.5cm]
-3&0  &1/2&-1/2&2&$(\nu ,e)_L$         \\[0.5cm]
+4&1/3&0  &+2/3&3&$u_R       $         \\[0.5cm]
-6&0  &0  &-1  &1&$e_R       $         \\[0.25cm]
\hline
\end{tabular*}
\end{center}
\vspace{0.3cm}

Note that we have obtained both quark- and lepton-like monopoles in our
model. In fact, if one were to try to only get quarks - as I tried to do for 
a long time - the scheme would fail since the winding that leads to monopoles 
is essentially around the ${\tilde U}(1)$ circle. For quark-like monopoles 
the absolute values of the ${\tilde U}(1)$ charge are 1/6, 2/6 and 4/6.
In any model where one gets these charges, one must necessarily also get
3/6 and 6/6 which would be the lepton-like monopoles.

\section{Confinement}

Now to complete the analogy with the dual superconductor of QCD, a color
confinement scheme needs to be found. This is easily achieved if we let
$$
{\tilde {SU}} (3) \rightarrow Z_3 \ .
$$
Then the colored monopoles get confined by $Z_3$ strings and all confined
clusters of monopoles are color singlets.

There are two points I would like to raise here. 
The first point is a response
to a question I have often been asked - if the ${\tilde {SU}}(3)$ is broken,
where are the massless gluons? There are two possible resolutions to
this puzzle. First, the gluons could
be some complicated string state which is massless, just like the
chromoelectric string in QCD is supposed to be some complicated 
(coherent) state of gluons. Second, on scales smaller
than the width of the $Z_3$ flux tubes formed in the ${\tilde {SU}} (3)$
breaking, the ${\tilde {SU}} (3)$ gauge particles can act as if they are 
massless. 

The second point that I want to raise is probably due to my lack of
understanding of the dual superconductor picture of QCD \cite{th}.
In this picture, confinement is explained as being due to two
kinds of $U(1)$ strings which do not interact. This does not seem quite
right to me since one would expect chromoelectric flux tubes to
interact with each other.
In the dual standard model picture, however, the confining strings 
are $Z_3$ strings and these do interact with each other - 
a string may terminate
at a vertex with two other strings (forming a Y junction).

Having recovered the magnetic
version of the charge spectrum of the standard
model and confinement, further questions arise: how can one get three
families of monopoles? why are the monopoles fermionic? what are their
masses? I will now describe some ideas on these issues.

\section{Families}

First consider the question of families. Topologically there should be
no difference between the dual electron ($\tilde e$) and the dual
muon ($\tilde \mu$) since the electric charges on the electron and the
muon are identical. So the difference between $\tilde e$ and $\tilde \mu$
must be non-topological, that is, dynamical. Could it be that there are
exactly two excited states of $\tilde e$ that can be identified with
$\tilde \mu$ and $\tilde \tau$? There is one scheme in which this is
precisely what happens \cite{hlgstv}. However, the scheme requires 
significant
extension of the ${\tilde {SU}}(5)$ model. Consider the breaking of a
simply connected symmetry group $G$ as follows:
\begin{eqnarray*}
G &\rightarrow& K \equiv 
[ \tilde{SU}_0(5)\times (\tilde{SU}(5))^3 ]/ (Z_5)^3 \\ 
  &\rightarrow& K' \equiv 
[\tilde{SU}(3) \times \tilde{SU}(2) \times \tilde{U}(1)] / 
                (Z_3 \times Z_2) \ .\\
\end{eqnarray*}
In this symmetry breaking, $K'$ is assumed to be a subgroup of 
$\tilde{SU}_0 (5)$ and the 
$(\tilde{SU}(5))^3$ breaks down to $(Z_5)^3$. Now the
monopoles in the final stage are given by the first homotopy group of 
$K'$ which is the same as in the ${\tilde {SU}}(5)$ model. However,
because of the intermediate stage, such monopoles are unstable to
fragmenting into smaller monopoles (``digits'') that would be
produced in the first stage of symmetry breaking. The digits are
confined by $Z_5$ strings produced in the $(\tilde{SU}(5))^3$ breaking.
So basically the $\tilde{SU}_0(5)$ monopoles have excitations that 
correspond to which of the three kinds of $Z_5$ strings are present.
Since there are three kinds of $Z_5$ strings in the model, there
are three families of monopoles.

The above scheme for obtaining three families is fundamentally quite
simple: we have provided three varieties of excitations of the 
${\tilde {SU}}(5)$ monopoles - one per (extra) 
$\tilde{SU}(5)$ factor - and so
the model contains three monopole families. At the same time, I must 
admit that I wish there was a simpler implementation. For example,
the present scheme would have to have an extremely complicated Higgs
structure which would be very hard to analyse and work with. 

One lesson that we learn from our effort to construct a model with
three families is that, from the perspective of the dual standard
model, the quarks and leptons must have internal excitations.
Further, in the only scheme that we know to get three families, it seems 
that the quarks and leptons must be composite and the underlying layer of
structure is in the spectrum of digits.

\section{Fermions}

There are at least two schemes by which monopoles can be made fermionic.
The first scheme is by introducing fermions in the original 
model \footnote{Introducing supersymmetry would be equivalent if somewhat 
more elaborate.}. Here, however, I want to discuss a scheme which is
closer to the spirit of the Sine-Gordon and Thirring model equivalence:
I would like to construct a purely bosonic dual standard model in which the 
monopoles can be fermionic. 
For 't Hooft-Polyakov \cite{thmono,apmono} monopoles, as a consequence of
an idea termed ``spin from isospin'' \cite{jr,hh}, it has been shown that
dyons in purely bosonic theories can carry half-integral spin,
from which it also follows that the dyons are fermionic \cite{ag}.
The scheme is to introduce a scalar in the fundamental representation of 
$SU(2)$. Bound states of this scalar with monopoles yield dyons of
spin 1/2. The scheme has also been generalized for $SU(5)$ monopoles with
unit winding \cite{ls}. We can easily extend the analysis to all the
${\tilde {SU}}(5)$ monopoles \cite{tvspin}.

The magnetic ${\tilde {SU}}(3)$ (generators $\lambda_3$ and $\lambda_8$), 
${\tilde {SU}}(2)$ (generator $\tau_3$), and, ${\tilde {U}}(1)$ (generator
$Y$) charges on a winding $n$ monopole may be written as:
$$
m_1 = 0\ ,\ \  m_2 = {{n_8} \over {\sqrt{3} g}}\ , \ \
m_3 = {{n_3} \over {2 g}}\ , \ \
m_4 = {{-1} \over {2g}} \sqrt{5 \over 3} n_1 \ ,
$$
where $g$ is the gauge coupling and
$$
n_8 = n +3k \ , \ \ n_3 = n + 2 l\ ,
\ \  n_1 = n \ ,
$$
where $k$ and $l$ can be arbitrary integers. However, for a 
particular choice
of $k$ and $l$ the numbers $n_8$ and $n_3$ are minimal and these
minimal values are often convenient to work with.

The electric charges on the various components of the 
${\tilde {SU}}(5)$ fundamental field $H$ are:
\begin{eqnarray*}
e_1^h &=& {g \over 2}\pmatrix{1\cr -1\cr 0\cr 0\cr 0\cr}\ , \
e_2^h = {{g} \over {2\sqrt{3}}}\pmatrix{1\cr 1\cr -2\cr 0\cr 0\cr}\ , \\
e_3^h &=& {{g} \over {2}}\pmatrix{0\cr 0\cr 0\cr 1\cr -1\cr}\ , \
e_4^h = {{g} \over {\sqrt{15}}}\pmatrix{1\cr 1\cr 1\cr -3/2\cr -3/2\cr}\ , \\
\end{eqnarray*}
where $h=1,...,5$ labels the components of $H$.
The angular momentum of a dyon is simply 
$$
J^h = - \sum_i m_i e_i^h
$$
and this leads to:
$$
J^h = \pmatrix{(- n_8+n_1)/6\cr (-n_8+n_1)/6\cr (+2n_8+n_1)/6\cr
                        (-n_3-n_1)/4\cr (+n_3-n_1)/4\cr}\ .
$$
Now it is easily checked that the stable monopoles can have half-integer
spin provided the electric charge due to $H$ is chosen to be in a suitable 
orientation, {\it i.e.} $J$ will be 
half-integral for some components 
of $H$ \footnote{In fact, half-integral 
spin can occur even in the absence of magnetic charge \cite{tvspin}.}.

In this way it is possible to get spin 1/2 dyons. However, note that
in addition to the spin 1/2 dyons, there are also dyons having
spin 0, 1, etc. That is, there is spin degeneracy (somewhat like
in supersymmetry). This might be acceptable if the spin 1/2 dyons
are the lightest states. How might this be?
As described in \cite{hltv}, a $\theta$ term offers
the possibility of reducing the electric charge on a dyon, thus making
it lighter, while keeping the spin and statistics unchanged. Also, a
$\theta$ term may be essential for eventually dualizing the model
since the quarks and leptons are known to be purely electric and 
not dyonic. These ideas have yet to be tried out in any detail and
I do not know if they can be made to work.

\section{Conclusions}

I would like to end this talk by summarizing the encouraging aspects
of this endeavour, the open problems and the predictions that are
indicated at this stage.

\smallskip

First the encouraging aspects:

$\bullet$ There is a one to one correspondence of the stable magnetic 
monopole charges with the electric charges of standard model fermions. 
That is,
the occurrence of the quarks and leptons is topological together with
a little dynamical input (in the stability argument).

$\bullet$ Monopole families are possible to construct and the
number of families is associated with a symmetry structure.

$\bullet$ The dualized monopoles occur in fundamental representations of 
$SU(3)$ and $SU(2)$. Here I am using the Goddard-Nyuts-Olive (GNO) 
conjecture \cite{gno}.
However, from another viewpoint, this is the only representation the
monopoles can come in because, by the Brandt-Neri-Coleman (BNC) 
analysis \cite{bnc,scstable},
monopoles that would occur in higher representations would be unstable.
An example is the diquark discussed in \cite{hltv}. This is a winding
two monopole in which the two monopoles have parallel ${\tilde {SU}}(3)$
orientation. According to the GNO conjecture, this would have to be
a 6 of $SU(3)$. But the diquark is unstable due to the BNC instability
and is absent from the spectrum.

$\bullet$ Color confinement is easy to see and quark-like monopoles are 
confined in color singlets.

$\bullet$ There is a scheme by which the monopoles can be made fermionic.

\smallskip

Next, some open issues:

$\circ$ How might one obtain chiral monopoles? In Table I, for example, 
what is
the analog of the $L$ and $R$ subscripts on the magnetic monopoles?

$\circ$ How can one make the monopoles have spin 1/2 and also perform
a duality rotation so that the dyons rotate into purely electric charges? 

$\circ$ How can one calculate the masses of the monopoles at strong
coupling? (This is also tied to the issue of chirality and to the
issue of electroweak symmetry breaking discussed below.)
The masses of monopoles can
be calculated at weak coupling, but we are interested in the strong
coupling regime where quantum corrections are extremely important.
Such a calculation is, in principle, allowed in the model but the
technology is not available. This is to be contrasted with the
standard model where the masses are not calculable even in principle
and the Yukawa couplings are fed in by hand.

$\circ$ The ${\tilde {SU}}(2) \times {\tilde U}(1)$ symmetry needs to
be broken down to ${\tilde U}(1)_{em}$ but a Higgs mechanism has the
undesirable consequence of connecting $\tilde e$, for example, to a
string. However, the ${\tilde {SU}}(2) \times {\tilde U}(1)$ is
strongly coupled and it seems reasonable to assume that the symmetry
could be broken in a way that magnetic fields are screened rather
than confined. In fact, the strongly coupled electroweak model has
been studied before \cite{ef} and the spectrum of states is found
to be identical to the spectrum of states in the weakly coupled
theory. This gives us hope that the spectrum of monopoles in the
strongly coupled ${\tilde {SU}}(2) \times {\tilde U}(1)$ is the
same as that if ${\tilde {SU}}(2) \times {\tilde U}(1)$ was weakly
coupled, {\it i.e.} exactly the states in Table I. 

$\circ$ If there are spin degeneracies in the monopole spectrum, why
are the spin 1/2 states the lightest?

\smallskip

Finally, at a speculative level, I would like to describe some of the
predictions that seem to be indicated even at this preliminary stage:

$\Diamond$ The electric charge spectrum of the standard model particles is
fixed - no values of hypercharge other than the ones presently 
observed are allowed.

$\Diamond$ The standard model particles have internal structure and this
is responsible for the family structure. If the scheme to obtain 
families described here is correct, quarks and leptons would be
composite.

$\Diamond$ If the ``spin from isospin'' scheme is correct, we expect to
see spin degeneracies in quarks and leptons. The quarks and leptons
will have scalar
partners but they need not be in supersymmetric multiplets. It
is also likely that some of these partners will be dyonic.

$\Diamond$ At some high energy in scattering experiments, it should become
clear that the known particles are really monopoles since particles and
solitons are known to scatter very differently.

\

{\it Acknowledgments:}
I would like to thank the organizers for holding such a stimulating
workshop. This work was supported by the Department of Energy.

\end{document}